\documentclass[
showpacs,preprintnumbers,amssymb]{revtex4}
\usepackage{amssymb}

\begin{document}

\tolerance=5000

\def\be{\begin{equation}}
\def\ee{\end{equation}}
\def\bea{\begin{eqnarray}}
\def\eea{\end{eqnarray}}
\def\tr{{\rm tr}\, }
\def\nn{\nonumber \\}
\def\e{{\rm e}}

\title{Scalar-tensor theory with Lagrange multipliers: a way of understanding the cosmological constant problem, and future singularities}

\author{Diego S\'aez-G\'omez\footnote{E-mail: saez@ieec.uab.es} }
\affiliation{Institut de Ci\`encies de l'Espai (ICE-CSIC/IEEC),\\
Campus  UAB, Facultat de Ci\`encies,  Torre C5-Parell-2a planta, 08193 Bellaterra (Barcelona) Spain, EU}
\begin{abstract}
The use of Lagrange multipliers in the context of quintessence/phantom scalar fields allows  to constrain the behavior of the scalar field, which provides a powerful tool, not only for the reconstruction of cosmological solutions but also for the study of some problems in cosmology and gravitational physics. In the present paper, we focus on the reconstruction of cosmological solutions capable of controlling the cosmological constant value by imposing a constraint on the scalar field,  providing a  relaxation mechanism of the value of the cosmological constant.  The formalism is also extended to the study of phantom scalar fields with a future singularity and their conformal transformation to the Jordan frame, where a type of  modified gravity, constrained by the Lagrange multiplier, is obtained.
\end{abstract}
\maketitle

\section{Introduction}

In the last decade, a large number of  Dark Energy candidates have been proposed in order to achieve explain the observational data, and also solve some old theoretical problems as the cosmological constant (cc) problem. However, the full understanding of the nature of Dark Energy is still very far of being complete. Firstly because of the impossibility to distinguish between the diverse theoretical models and the absence of new predictions, but also because of the complexity of some of these models, as modified gravity (for  recent reviews on modified gravity and the reconstruction of cosmological solutions, see \cite{Nojiri:2010wj}). Nevertheless, it was proposed in \cite{Lagrange2} a model which might provide the necessary tools for a better understanding of some of the existing problems in  cosmological models and in gravitational physics. The model consists of a common quintessence/phantom scalar field, but as a novelty, the evolution of the scalar field has to satisfy a constraint introduced  through a Lagrange multiplier, which basically behaves as an additional non-propagating scalar field, that retains a single dynamical degree of freedom. By a particular choice of the constraint, the differential equations for the two scalar fields, the quintessence/phantom field and the Lagrange multiplier, become first order equations, so that there are no propagating wave-like degrees of freedom.  Such formalism can be easily extended to modified gravity instead of a scalar field, as it was proved in Ref.~\cite{Lagrange1}, which also provides interesting properties. This approach has  shown that  cosmological solutions which reproduce dark energy epoch can be easily reconstructed, but also it provides the necessary tools to solve the problems of a particular model, as violations at local scales or  additional propagating modes of the graviton, which can be avoided  by the constraint  of the Lagrange multiplier   (see \cite{Lagrange2}-\cite{Gao:2010gj}). Hence, the formalism can be used not only as a tool to reconstruct solutions but also to identify and isolate intrinsic problems of a particular model. \\

The main purpose of this paper is the reconstruction of models capable to reproduce late-time acceleration,  providing a natural explanation of why the value of the observed cosmological constant  (cc) at the present time is so small compared with the vacuum energy density predicted by quantum field theories. By  using the formalism of a scalar field constrained by a Lagrange multiplier,  a relaxation mechanism of the value of the cc can be constructed, so that the cc problem can be dealt through the constraint on the scalar field, and a better understanding on its the evolution can be obtained (for other kind of relaxation mechanisms and their evolution, see \cite{Sola}). Even, by such mechanism  the inflationary epoch can be also reproduced, providing a unified picture of the entire Universe evolution. These results can be easily extended in terms purely of the metric tensor, as it was shown in Ref.~\cite{Lagrange1}, which may provide a more natural explanation, where the gravitational field is subject to some constraints. \\

On the other hand, such model allows to study the behavior of some future singularities when a conformal transformation is applied, which relates the Einstein frame and the Jordan one. Specifically, we focus on the reconstruction of $F(R)$ gravity from a phantom Universe described in the Einstein frame. Phantom cosmological models have been widely studied, as they carry very interesting physical information (see Refs.~\cite{phantom}-\cite{Scalar-tensor}). It is well known that $F(R)$ gravity is equivalent to a scalar-tensor theories class with a non-minimally coupling scalar field and  null kinetic term, such that the action in the Einstein frame with a scalar field can be transformed into a $f(R)$ action through a particular conformal transformation (see Refs.~\cite{Barrow:1988xh,Maeda:1988ab,F(R)fromScalar}). However, this is restricted to the cases of a quintessence field, free of future singularities (see Refs.~\cite{F(R)vsScalar1}), while for the phantom case, the action becomes complex after being conformally transformed (see Ref.~\cite{F(R)vsScalar2}), unless an additional component is included \cite{Saez-Gomez1}. By using  a Lagrange multiplier,  a type of $F(R)$ gravity can be obtained in the Jordan frame starting from a phantom solution in the Einstein frame. Hence, the conformal transformation of future singularities (for a classification of future singularities see Ref.~\cite{Nojiri:2005sx}), specifically the Big Rip singularity,  is  studied, which provides a better understanding of the future singularities when a particular conformal transformation is applied. \\

The paper is organized as follows: in the next section, we introduce the quintessence/phantom models with a Lagrange multiplier. Then, in section III we reconstruct cosmological solutions with a relaxing mechanism for the cosmological constant. In section IV, we reconstruct $F(R)$ gravity in the Jordan frame starting from quintessence/phantom scalar fields  in the Einstein frame. Finally, in the last section we analyze the transformation of a phantom field with a Big Rip singularity into $f(R)$ gravity in the Jordan frame.

\section{Scalar-tensor theory with Lagrange multiplier}

In this section, we introduce the model where a minimally coupled scalar field is considered with the presence of  a Lagrange multiplier that imposes a constraint on the scalar field. In the next sections, the reconstruction of cosmological solutions is done, where the Lagrange multiplier plays an important role. The general action considered here is given by,
\be
S=\int d^4x \sqrt{-g}\left\{ \frac{R}{2\kappa^2}-\frac{\omega(\phi)}{2}\partial_{\mu}\phi\partial^{\mu}\phi-V(\phi)-\lambda F(\phi, \partial_{\mu}\phi\partial^{\mu}\phi,...)+\mathcal{L}_m\right\}\  .
\label{1.1}
\ee
Here $\kappa^2=8\pi G$ and $\lambda$ is the Lagrange multiplier, while $\mathcal{L}_m$ refers to the matter lagrangian. The function $F(\phi, \partial_{\mu}\phi\partial^{\mu}\phi,...)$ is the constraint on the scalar field $\phi$, while $\omega(\phi)$ and $V(\phi)$ are the kinetic term and the scalar potential respectively. Note that  the Lagrange multiplier behaves as a second scalar field, which does not propagate. By varying the action (\ref{1.1}) with respect the metric $g_{\mu\nu}$, the field equations are obtained,
\[
R_{\mu\nu}-\frac{1}{2}g_{\mu\nu}R=\kappa^2\left(T_{\mu\nu}^{(m)}+T_{\mu\nu}^{(\phi,\lambda)}\right)\ , 
\]
where,
\[
T_{\mu\nu}^{(\phi,\lambda)}=\omega(\phi)\partial_{\mu}\phi\partial_{\nu}\phi-g_{\mu\nu}\left(\frac{\omega(\phi)}{2}\partial_{\mu}\phi\partial^{\mu}\phi+V(\phi)+\lambda F(\phi, \partial_{\mu}\phi\partial^{\mu}\phi,...)\right)+2\lambda\frac{\delta F(\phi, \partial_{\mu}\phi\partial^{\mu}\phi,...)}{\delta g_{\mu\nu}}\ , \nn
\]
\be
T_{\mu\nu}^{(m)}=-\frac{2}{\sqrt{-g}}\frac{\delta\mathcal{L}_m}{\delta g_{\mu\nu}}\ .
\label{1.2}
\ee
While the variation over $\lambda$ gives the constraint equation on the scalar field $\phi$,
\be
F(\phi, \partial_{\mu}\phi\partial^{\mu}\phi,...)=0\ .
\label{1.3}
\ee
Hence, solving  the  equation (\ref{1.3}), the evolution of the scalar field $\phi$ is fixed, which affects  the Einstein field equations (\ref{1.2}).  For some specific functions $F$,  one can obtain interesting properties and cosmological solutions, where a relaxation mechanism of the vacuum energy density can be implemented, as shown below. In this paper, we are interested in studying spatially flat FLRW, whose metric is given by,
\be
ds^2=-dt^2+a(t)^2\sum_{i=1}^3 dx^{i 2} \ .
\label{1.4}
\ee
Then, by introducing the metric (\ref{1.4}) in the field equations (\ref{1.2}) and specifying the function $F(\phi, \partial_{\mu}\phi\partial^{\mu}\phi,...)$, the usual FLRW equations are obtained. Let us consider a simple example in order to show the effects of the constraint equation (\ref{1.3}),
\be
F(\phi, \partial_{\mu}\phi\partial^{\mu}\phi,...)=F(\phi)=0\ ,
\label{1.5}
\ee
By solving this equation, the solution gives a constant scalar field $\phi(t)=\phi_0$,  the roots of the equation (\ref{1.5}). Hence, the energy-momentum tensor gives  an effective cosmological constant $T_{\mu\nu}^{(\phi,\lambda)}=-g_{\mu\nu}V_0$, which can compensate for the large contribution of the vacuum energy density. Note that in general, equation (\ref{1.5}) contains more than one positive root $\phi_0$, so that each point provides an effective cosmological constant which may reproduce, or at least contribute, to the periods of accelerating expansion of the Universe history, unifying inflation and late-time acceleration under the same mechanism. However, here we are interested on those cases with a non-constant scalar field as more interesting results may be obtained, so from now on we consider the constraint equation,
\be
F(\phi, \partial_{\mu}\phi\partial^{\mu}\phi)=\frac{1}{2}\partial_{\mu}\phi\partial^{\mu}\phi + U(\phi)=0\ .
\label{1.6}
\ee
As we are considering an homogeneous and isotropic metric such as FLRW, the scalar field is taken only as a function of time dependent $\phi=\phi(t)$, so that the constraint (\ref{1.6}) on the scalar field is,
\be
\frac{1}{2}\dot{\phi}^2-U(\phi)=0 \quad \rightarrow \quad \frac{d\phi}{\sqrt{2U(\phi)}}=\pm dt\ .
\label{1.8}
\ee
By the metric (\ref{1.4}), and using the condition (\ref{1.8}), the FLRW equations are given by,
\[
H^2=\frac{\kappa^2}{3}\left[\left(\omega(\phi)+2\lambda\right)U(\phi)+V(\phi)+\sum_{i}\rho_{i} \right]\ ,
\]
\be
-3H^2-2\dot{H}=\kappa^2\left[\omega(\phi)U(\phi)-V(\phi)+\sum_{i}w_{i}\rho_{i}\right]\ .
\label{1.7}
\ee
Here, we are considering matter species described by perfect fluids, where the subscript $i$ refers to every kind of matter in the Universe with an equation of state  given by $p_i=w_i\rho_i$. Then, by using the above equations, the corresponding cosmological solution is obtained. The evolution of every matter specie can be easily calculated by solving the continuity equation,
\be
\dot{\rho_i}+3H(1+w_i)\rho_i=0 \quad \rightarrow \quad \rho_i(t)=\rho_{0i}\e^{-3(1+w_i)\int^t dt'H(t')}\ ,
\label{1.9}
\ee
where $\rho_{0i}$ is an integration constant. Then, we can give a reconstruction method, similar to the case of free scalar fields (Ref. \cite{Scalar-tensor}), in order  to reconstruct the corresponding theory for  some particular solutions $H(t)$ and $\phi(t)$. As the  expression for $\phi$ can be inverted,  the time coordinate is written as a function of the scalar field $t=t(\phi)$, and by using the constraint equation (\ref{1.8}) and  the FLRW equations (\ref{1.7}), the theory that reproduce a specific solution is described by,
\be
U(\phi)=\frac{\dot{\phi}(t(\phi))^2}{2}\ , \quad V(\phi)=\frac{1}{\kappa^2}\left[3H(t(\phi))^2+2\dot{H}(t(\phi))\right]+\omega U(\phi)+\sum_iw_i\rho_{0i}\e^{-3(1+w_i)\int^{t(\phi)} dt'H(t')}\ ,
\label{1.10}
\ee
while the Lagrange multiplier yields,
\be
\lambda(t)=-\frac{1}{2U(\phi(t))}\left[2\omega U(\phi(t))+\frac{2}{\kappa^2}\dot{H}(t)+\sum_i(1+w_i)\rho_{0i}\e^{-3(1+w_i)\int^t dt'H(t')}\right]\ .
\label{1.11}
\ee
Note that $\omega(\phi)$ in the action (\ref{1.1}) remains arbitrary, and we have assumed here a constant kinetic term $\omega(\phi)=\omega$. Hence, we have implemented a method to reconstruct the corresponding  action (\ref{1.1}) for any given solution. Let us consider the example described by the solutions,
\be
\phi(t)=\phi_0t\ , \quad H(t)=H_0+\frac{H_1}{t}\ ,
\label{1.12}
\ee
where $H_0$, $H_1$, and $\phi_0$ are constants. This solution describes a Universe that crosses through out two phases, the first one dominated by the second term in r.h.s. of the Hubble expression (\ref{1.12}), which produces a phase of decelerating expansion, identified with the radiation/matter dominated epochs. Then, a second phase dominated by the constant $H_0$, which describes a de Sitter solution at late times is reproduced. By using the expressions (\ref{1.10}), the corresponding  potentials for the scalar field are obtained,
\be
U(\phi)=\frac{\phi_0^2}{2}\ , \quad V(\phi)=\frac{\omega\phi_0^2}{2}+\frac{1}{\kappa^2}\left[3H_0^2+\frac{6H_0H_1\phi_0}{\phi}+\frac{H_1\phi_0(3H_1\phi_0-2)}{\phi^2}\right] +\sum_iw_i\rho_{0i} \left(\frac{\phi}{\phi_0}\right)^{-3H_1(1+w_i)}\e^{\frac{-3(1+w_i)H_0}{\phi_0}\phi}\ .
\label{1.13}
\ee
While the Lagrange multiplier (\ref{1.11}) for this particular case is given by,
\be
\lambda(t)=-\frac{1}{\phi_0^2}\left[\omega\phi_0^2-\frac{2H_1\phi_0}{\kappa^2}\frac{1}{t^2}+\sum_i(1+w_i)\rho_{0i} t^{-3H_1(1+w_i)}\e^{-3(1+w_i)H_0t}\right]\ .
\label{1.14}
\ee
Hence, we have reconstructed a theory, described by the action (\ref{1.1}) and the potentials  in (\ref{1.13}), which reproduces the solution (\ref{1.12}). The behavior of the Lagrange multiplier is also fixed by (\ref{1.14}). Note that any kind of  matter specie can be included in the equations (\ref{1.13}). We might include the vacuum energy density, whose EoS is given by the same one of a cosmological constant, i.e. $p_{vac}=-\rho_{vac}$, and its contribution to the dark energy density would be compensated by the presence of the scalar field. \\
By this simple reconstruction method, we have shown that the constraint imposed on the evolution of the scalar field (\ref{1.8})  affects the expansion history and by a suitable choice of the potentials (\ref{1.10}), the large expected value of the vacuum energy density can be relaxed by a natural mechanism. The next section is devoted to the analysis of  some examples in  more detail, where also the unification of inflation and dark energy is considered.

\section{Relaxing the cosmological constant}

We are interested here to study the  relaxation mechanism of the vacuum energy density in detail with explicit examples that are able to reproduce the dark energy epoch, and even the inflationary era.  In the previous section, a general reconstruction method of cosmological solutions was presented, and some examples were studied. Nevertheless, here we are more interested in studying the evolution of the vacuum energy density, and the  scalar field contribution to the dark energy density, and how the implementation of a relaxation mechanism can be realized by using the Lagrange multiplier. Let us consider the FLRW Universe described by the second equation in (\ref{1.7}),
\be
-\frac{1}{\kappa^2}(3H^2+2\dot{H})=w_m \rho_m-\rho_{vac}+\frac{1}{2}\dot{\phi}^2-V(\phi)\, ,
\label{2.1}
\ee
here we have used the constraint equation (\ref{1.8}), $U(\phi)=\frac{1}{2}\dot{\phi}$. We have also included the energy density for matter $\rho_m$, the vacuum energy density, $\rho_{vac}=\frac{\Lambda}{\kappa^2}$, and we have assumed  a constant kinetic term, given by $\omega(\phi)=1$. We  define an effective cosmological constant in the equation (\ref{2.1}) that  depends on time, and which can be written as,
\be
\Lambda_{eff}(t)= \Lambda-\kappa^2\left(\frac{1}{2}\dot{\phi}^2-V(\phi)\right)= \Lambda-\kappa^2\left(U(\phi)-V(\phi)\right)\ . 
\label{2.2}
\ee
By the appropriate scalar potentials, we can implement a relaxation mechanism of the vacuum energy density. Then, the FLRW equation (\ref{2.1}) takes the form,
\be
-3H^2-2\dot{H}=\kappa^2w_m \rho_m-\Lambda_{eff}\, ,
\label{2.1a}
\ee
Hence, the terms depending on $\phi$ in (\ref{2.2}) may act by relaxing the cosmological constant $\Lambda$, such that the effective value  coincides with the one observed, $\Lambda_{eff}(t_0)\sim H_0^2$, where the subscript $_0$ refers to the present epoch. In this sense, we will give some examples, starting with  simple models where there is only a constant contribution to the effective value of the cosmological constant,  to more complex models that are  also capable of unifying the inflationary and dark energy eras.

\subsection{Example 1}

Let us consider a model where the potential $U(\phi)$ is given by,
\be
U(\phi)=\frac{\Lambda_{\phi}}{\kappa^2}+V(\phi)\, ,
\label{2.3}
\ee
where $\Lambda_{\phi}$ is a positive constant. Then, the effective cosmological constant (\ref{2.2}) yields,
\be
\Lambda_{eff}=\Lambda-\Lambda_{\phi}\ .
\label{2.4}
\ee
Hence, the value of the cosmological constant is compensated by the contribution of the scalar field. And in this case, the equation for the Hubble parameter (\ref{2.1a}) turns out to be the same as for the $\Lambda$CDM model, whose solution can be easily calculated, 
\be
H(t)= \sqrt{\frac{\Lambda_{eff}}{3}} \coth\left( \frac{(1+w_m)\sqrt{3\Lambda_{eff}}}{2} t\right)\ .
\label{2.5}
\ee
While the evolution of the scalar field can be calculated by introducing the potential (\ref{2.3}) in the equation (\ref{1.8}), and by specifying the potential $V(\phi)$. If we consider for this case, $V(\phi)=\frac{1}{2}m^2\phi^2$ with $m$ being a constant,   the solution for the scalar field is given by,
\be
\phi(t)=\sqrt{\frac{2}{\kappa^2}\frac{\Lambda_{\phi}}{m^2}}\sinh\left(m\sqrt{\frac{1}{2}} t\right)\ .
\label{2.6}
 \ee
On the other hand, the expression for the Lagrange multiplier  can be easily calculated by introducing the above quantities in the first FLRW equation in (\ref{1.7}),  which yields $\lambda(t)$. Hence, this simple model shows a way to relax the value of the vacuum energy density by means of  the scalar field, and the corresponding Lagrange multiplier. The  evolution is the same as in the $\Lambda$CDM model,  and the effective cosmological constant can be adjusted with the observational data, setting the value of $\Lambda_{\phi}$. In the next examples, we study more complex models, where the effective cosmological constant can change with time, and the unification of the inflationary and dark energy epochs can be realized.

\subsection{Example 2}
We are now interested to consider a model where the effective cosmological constant (\ref{2.2})  is time-dependent. Let us consider now the potentials, 
\be
U(\phi)=2V(\phi)=\frac{1}{2}m^2\phi^2\ ,
\label{2.7}
\ee
where $m$ is a constant. It is straightforward to calculate the evolution of the scalar field  by solving the constraint equation (\ref{1.8}), which gives,
\be
\phi(t)=\phi_0\e^{-\frac{\alpha}{2}t}\ .
\label{2.8}
\ee
Here $\alpha^2=8m^2$. Then, for this model,  the effective cosmological constant (\ref{2.2}) is now given by,
\be
\Lambda_{eff}= \Lambda-\Lambda_{\phi}\e^{-\alpha t}\ ,
\label{2.9}
\ee
where $\Lambda_{\phi}=\frac{\kappa^2\alpha^2}{16}\phi_0^2$, and we assume $\Lambda_{\phi}<\Lambda$. In this case, the effective cosmological constant evolves with  time, whose behavior will depend on the value of $\alpha$ that determines the growth rate of the effective cc. In this model, the general solution of the equation (\ref{2.1a}) is given by a much more complex function for the Hubble parameter than in the model analyzed above. However, for short time intervals, the solution can be well approximated by the $\Lambda$CDM solution, given in (\ref{2.5}). Note that the minimum value of the cc (\ref{2.9}) is given at $t=0$, and the radiation/matter epochs can be reproduced.  Then, the cc constant grows and the Universe enters in the dark energy epoch, as one expects. Note also that the value of $\alpha$ has to be very small in units of $Gyrs^{-1}$ in order to match the observational value  $\rho_{\Lambda_{eff}}\sim 10^{-47}\ GeV^{4}$ at the current epoch.

\subsection{Example 3}

As a last example, we want to consider a model capable to unify inflation and dark energy epoch under the same mechanism, assuming that the vacuum energy density dominates during a short time of period right after the origin of the Universe providing a large value of the cosmological constant, then it decreases down a minimum limit until the present epoch, where it  increases again. Hence, the $U$ and $V$ potentials considered here are,
\be
U(\phi)=\frac{\phi^2}{2}\ , \quad V(\phi)=\frac{\phi_0^2\kappa^2(\alpha\phi^2+\phi_0^2)-2\Lambda_{\phi}\phi_0(\phi-\phi_0 t_0)}{2\kappa^2(\alpha\phi^2-\phi_0^2)}\ .
\ee
Here $\Lambda_{\phi}$, $\phi_0$, $\alpha$ and $t_0$ are constants. Then, by the equation (\ref{1.8}), the evolution of the scalar field is given by,
\be
\phi(t)=\phi_0 t\ .
\label{2.10}
\ee
While the effective cosmological constant is now described by,
\be
\Lambda_{eff}= \Lambda-\Lambda_{\phi}\frac{t-t_0}{\alpha t^2+1}\ .
\label{2.11}
\ee
This expression describes a cosmological constant that has a maximum value at $t=t_0$, where inflation occurs caused by the vacuum energy density, then the value of the effective cc decreases to a minimum, after which it starts to grow again. Hence,  this model is able to reproduce the unification of the accelerating epochs of the Universe. It is straightforward to show that the minimum of the function (\ref{2.11}) is given at $t_{min}=t_0+\sqrt{t_0^2+\frac{1}{\alpha}}$, so that if we assume $\Lambda_{\phi}\sim 2\Lambda\alpha\left(t_0+\sqrt{t_0^2+\frac{1}{\alpha}}\right)$, the minimum value of the effective cc (\ref{2.11}) can be fitted around zero at the present age of the Universe matching the small value given by the  observational  data. Following this example, the value of the effective cc should  increase in the future, the time it takes to observe the possible effects depends entirely on $\alpha$. Hence,  an example  that provides a relaxation mechanism for the value of the vacuum energy density at the present time but which also reproduces the early accelerated epoch, has been reconstructed. The results achieved here can be easily extended to modified gravities instead of the use of a scalar field, which can give a natural answer to the evolution of the Universe.

\section{Reconstructing modified gravity with Lagrange multipliers}

We have shown that the use of Lagrange multipliers can provide a useful reconstruction technique, where interesting properties on the cosmological paradigm can be obtained. Particularly,  the relaxation of the value of the cosmological constant can be easily achieved by using a scalar field  constrained by a Lagrange  multiplier. The same technique can be extended to  modified gravity, as it was shown in Ref.~\cite{Lagrange1}. \\
In this section we are interested to transform the action (\ref{1.1}) into the Jordan frame in order to obtain the corresponding $F(R)$ gravity. It is well known that $F(R)$ gravity is equivalent to a type of Brans-Dicke theory with a potential and a null kinetic term written in the so-called Jordan frame, such that $F(R)$ gravity can be obtained from quintessence-like scalar fields by applying a particular conformal transformation on the Einstein frame (see Refs.~\cite{F(R)fromScalar}-\cite{Saez-Gomez1}). Here we are interested to extend this conformal transformation to the case of a minimally coupled scalar field  in the presence of a Lagrange multiplier. By applying a particular conformal transformation on the action (\ref{1.1}), the corresponding $F(R)$ action is obtained \cite{F(R)fromScalar}, 
\be
g_{\mu\nu}=\Omega^2\tilde{g}_{\mu\nu}\, \quad \text{where} \quad \Omega^2=\exp\pm\kappa\int d\phi \sqrt{\frac{2}{3}\omega(\phi)}\ .
\label{3.1}
\ee
Here the ``tilde'' on the metric refers to the Jordan frame. Note that the transformation (\ref{3.1}) suppresses the kinetic term proportional to $\omega(\phi)$ in order to ensure that the corresponding Brans-Dicke-like action in the Jordan frame is equivalent to $F(R)$ gravity (see Refs.~\cite{Barrow:1988xh,Maeda:1988ab}). However,  the terms in front of the Lagrange multiplier remains also in the Jordan frame, where the action in vacuum is transformed as,
\be
S_{J}=\int dx^4 \sqrt{-\tilde{g}}\left[\frac{\Omega^2(\phi)}{2\kappa^2}\tilde{R}- \tilde{V}(\phi)-\lambda\left(\frac{\Omega^2(\phi)}{2}\tilde{\nabla}_{\mu}\phi\tilde{\nabla}^{\mu}\phi+\tilde{U}(\phi)\right)\right]\ ,
\label{3.2}
\ee
where $\tilde{V}(\phi)=\Omega^4(\phi) V(\phi)$ and $\tilde{U}(\phi)=\Omega^4(\phi) U(\phi)$. By varying the action (\ref{3.2}) with respect to $\lambda$,  the constraint equation is obtained, similarly to the case of the above sections. Note that in general, the action (\ref{3.2}) is not equivalent to any $F(R)$ gravity unless one imposes an additional condition on the conformal transformation (\ref{3.1}). By varying the action (\ref{3.2}) with respect  to the scalar field, the equation for  $\phi$ is obtained,
\be
\Omega(\phi)\frac{d\Omega(\phi)}{d\phi}\frac{\tilde{R}}{\kappa^2}-\frac{d\tilde{V}(\phi)}{d\phi}+\lambda\left[\Omega^2(\phi)\tilde{\Box}\phi+ \Omega(\phi)\frac{d\Omega(\phi)}{d\phi}\tilde{\nabla}_{\mu}\phi\tilde{\nabla}^{\mu}\phi-\frac{d\tilde{U}(\phi)}{d\phi}\right]=0
\label{3.3}
\ee
Then, in order to obtain $F(R)$ gravity in the Jordan frame,  the last  term  in (\ref{3.3}) have to become null, which can be achieved by imposing $\lambda=0$ that eliminates the constraint on the scalar field and the action in the Einstein frame (\ref{1.1}) reduces to the usual  action for a quintessence/phantom field, or we  can impose that terms inside the brackets are null, which implies the condition,   
\be
\tilde{\Box}\phi+\frac{1}{\Omega(\phi)}\frac{d\Omega(\phi)}{d\phi}\tilde{\nabla}_{\mu}\phi\tilde{\nabla}^{\mu}\phi- \frac{1}{\Omega^2(\phi)}\frac{d\tilde{U}(\phi)}{d\phi}=0\ .
\label{3.4}
\ee
This condition basically restricts the form of the kinetic term $\omega(\phi)$, which remained arbitrary in the Einstein frame as  showed in the previous sections. Here, it will be shown that $F(R)$ gravity can be obtained from phantom universes in the Einstein frame by applying the conformal transformation obtained, contrary to the usual phantom case (see \cite{F(R)vsScalar2}). Hence, by the condition (\ref{3.4}), the equation (\ref{3.3}) yields, 
\be
\Omega(\phi)\frac{d\Omega(\phi)}{d\phi}\frac{\tilde{R}}{\kappa^2}-\frac{d\tilde{V}(\phi)}{d\phi}=0 \quad \rightarrow \quad \phi=\phi(\tilde{R}) .
\label{3.5}
\ee
By solving this equation, we can obtain the corresponding action for $F(\tilde{R})$,
\be
S_J =\int dx^4 \sqrt{-\tilde{g}}\left[F(\tilde{R})+\lambda\left(\frac{W(\tilde{R})}{2}\tilde{\nabla}_{\mu}\tilde{R}\tilde{\nabla}^{\mu}\tilde{R}+U(\tilde{R})\right)\right]\ .
\label{3.6}
\ee
where,
\be
F(\tilde{R})=\frac{\Omega^2(\phi(\tilde{R}))}{2\kappa^2}\tilde{R}- \tilde{V}(\phi(\tilde{R}))\ , \quad \text{and} \quad W(\tilde{R})=\Omega^2(\phi(\tilde{R}))\left(\frac{d\phi(\tilde{R})}{d\tilde{R}}\right)^2\ .
\label{3.7}
\ee
The action (\ref{3.6}) corresponds to a type of  $F(R)$ gravity introduced first in \cite{Lagrange1}, where the scalar Ricci $\tilde{R}$, and for instance the metric tensor, is constrained by the Lagrange multiplier, similar to the case of scalar fields studied above. It is straightforward to show that assuming a flat FLRW Universe in the Jordan frame, a relaxation mechanism of the cc can be reconstructed starting from the action (\ref{3.7}),  but in this case, the relaxation is realized in terms of the metric instead of scalar fields, what can provide a more natural explanation to the observed value of the cosmological constant. However, here we are interested to show the mathematical relation between the action defined in the Einstein frame in terms of a scalar field (\ref{1.1}), and the $F(\tilde{R})$ action obtained in the Jordan frame (\ref{3.6}), and specially  the relation between the corresponding cosmological solutions. Note that we are referring here to the mathematical equivalence between both frames which may help to understand their physical relation, however, the physical equivalence of both frames  is a wide open discussion beyond the aim of this paper (for a review see Ref.~\cite{Faraoni:1998qx}). It is well known that phantom scalar fields defined in the Einstein frame has not a $F(R)$ correspondence in the Jordan frame (see Ref.~\cite{F(R)vsScalar2}), unless additional components are assumed (see Ref.~\cite{Saez-Gomez1}). However, in the case of the action (\ref{1.1}), the conformally transform to $F(R)$ gravity  is not restricted only to the non-phantom case. Let us first write the relation between the cosmological solutions for each frame. By assuming a flat FLRW metric in each frame, the metric components and coordinates in both frames are related by means of the conformal transformation (\ref{3.1}),
\be
\tilde{a}(\tilde{t})=\frac{a(\tilde{t})}{\Omega}\ , \quad d\tilde{t}=\frac{dt}{\Omega(\phi(t))}
\label{3.8}
\ee
Let us consider now an explicit example, where  the corresponding $F(R)$ action is obtained as well as  the cosmological solutions in the Jordan frame starting from  the Hubble parameter in the Einstein frame, 
\be
H=H(t)\ , \quad \phi(t)=\phi_0 t\ .
\label{3.9}
\ee
By the previous section  the solution for the scalar field (\ref{3.9}) corresponds to a constant potential $U(\phi)=\frac{\phi_0^2}{2}$, according to the constraint equation (\ref{1.8}).  We can now rewrite the equation (\ref{3.4}) in terms of the variables defined in the Einstein frame according to the transformations (\ref{3.8}),
\be
\Omega(\phi)\ddot{\phi}-\frac{d\Omega}{d\phi}\dot{\phi}^2+3\Omega H\dot{\phi}+4U(\phi)\frac{d\Omega}{d\phi}+\Omega\frac{dU(\phi)}{d\phi}=0\ . 
\label{3.10}
\ee
Here the dots on the scalar field $\phi$ refers to derivatives with respect the time $t$ defined in the Einstein frame. By assuming the solution (\ref{3.9}), the equation (\ref{3.10}) is considerably simplified,
\be
\phi_0(2-\phi_0)\frac{d\Omega}{d\phi}+3\phi_0H\Omega(\phi)=0\ .
\label{3.11}
\ee
By conformal transformation $\Omega$ defined in (\ref{3.1}), the equation (\ref{3.11}) can be solved for the kinetic term $\omega(\phi)$, which gives
\be
\omega(\phi)=\frac{27}{8\kappa^2(2-\phi_0)^2}H^2(t(\phi))\ .
\label{3.12}
\ee
Hence, by introducing this function in the conformal transformation (\ref{3.1}), we are able to obtain the corresponding $F(R)$ gravity in the Jordan frame (\ref{3.6}). Note also that for this explicit example, $\omega(\phi)>0$ independently of the  cosmological solution $H(t)$  in the Einstein frame, such that a phantom Universe can be conformally transformed in terms of a $F(R)$ action in the Jordan frame, which supposes  a difference with respect to the usual phantom scalar fields \cite{F(R)vsScalar2}. 

\section{Phantom fields and future singularity in the conformal frame}

Let us now analyze the transformation of an explicit Hubble parameter that contains a Big Rip singularity in the Einstein frame,
\be
H(t)\sim \frac{1}{t_s-t}\ \quad \rightarrow \quad a(t)=\frac{a_0}{t_s-t},
\label{3.13}
\ee
where $a_0$ is an integration constant and $t_s$ is the Rip time. The solution (\ref{3.13}) describes a super accelerating Universe in the Einstein frame that ends in the so-called Big Rip singularity at $t=t_s$. We are now interested to see  how the singularity is transformed  in the Jordan frame (\ref{3.6}). By introducing the expressions (\ref{3.9}) and(\ref{3.13}) in (\ref{3.12}), the kinetic term takes the form,
\be
\omega(\phi)= \frac{27}{8\kappa^2[(2-\phi_0)(t_s-\phi/\phi_0)]^2}\\ .
\label{3.14}
\ee
While the scalar potential $V(\phi)$ for the solution (\ref{3.13}) can be obtained by using the expression (\ref{1.10}), 
\be
V(\phi)=\frac{V_0}{(t_s-\phi/\phi_0)^2}\ ,
\label{3.14b}
\ee
where $V_0=\frac{160+\phi(107\phi-320)}{16(2-\phi_0)^2\kappa^2}$. The conformal transformation (\ref{3.1})  is  given by,
\be
\Omega^2(\phi)=\left(t_s-\frac{\phi}{\phi_0}\right)^{\mp \frac{3\phi_0}{2-\phi_0}}\ .
\label{3.15}
\ee
Hence, the cosmological solution (\ref{3.13}) is transformed in the Jordan frame through (\ref{3.8}), which yields
\be
\tilde{a}(\tilde{t})=a_0 \tilde{t}^{-1+\frac{4\phi_0-8}{\pm3\phi_0-4+2\phi_0}}\  \quad \rightarrow \quad \tilde{H}=\frac{1}{\tilde{a}}\frac{d\tilde{a}}{d\tilde{t}}\sim\pm\frac{1}{\tilde{t}}\ . 
\label{3.16}
\ee
where $a_0=\left(\mp\frac{2\phi_0-4}{3\phi_0}\right)^{-1+\frac{4\phi_0-8}{\pm3\phi_0-4+2\phi_0}}$. This solution describes an expanding  or contracting Universe depending on the sign in (\ref{3.14}) and on the value of the constant $\phi_0$. Nevertheless, independently of the type of the cosmological evolution, the solution (\ref{3.16}) contains a singularity at $\tilde{t}=0$ that can be identified as a Big Bang (Crunch). Therefore, the singularity in the Einstein frame (the Big Rip singularity) is  transformed in the Jordan frame as an initial singularity, which means that the conformal transformation (\ref{3.1}) does not remove the singular point but  displaces it.\\
In order to reconstruct the $F(R)$ action in the Jordan frame, one has to solve the equation (\ref{3.5}). For the example studied here, the solution is given by,
\be
\phi(\tilde{R})=\phi_0\left[t_s-(\alpha\tilde{R})^{\frac{\phi_0-2}{\pm3\phi_0+2(2-\phi_0)}}\right]\ \quad \text{where} \quad  \alpha=\pm\frac{3\phi_0}{4\kappa^2V_0(\pm3\phi_0+2-\phi_0)}\ .
\label{3.17}
\ee
And the $F(\tilde{R})$ action in the Jordan frame (\ref{3.6}) yields,
\be
 F(\tilde{R})=\left[\frac{(\alpha\tilde{R})^{1+\frac{2(\phi_0-2)}{\pm3\phi_0+4-2\phi_0}}}{2\kappa^2}-V_0\alpha\right]\tilde{R}\ , \quad  W(\tilde{R})=\beta(\alpha\tilde{R})^{\frac{6(\phi_0-2)\mp3\phi_0}{\pm3\phi_0+4-2\phi_0}}\ ,
 \label{3.18}
 \ee
where $\beta=\frac{9(\phi_0-2)^2\phi_0^4}{16\kappa^4V_0^2(\pm\phi_0+4-2\phi_0)^2(\pm\phi_0+2-\phi_0)^2}$. Hence, we have reconstructed the corresponding $F(\tilde{R})$ gravity in the Jordan frame, whose cosmological solution is given in (\ref{3.16}). Therefore, an action with a phantom scalar field constrained by a  Lagrange multiplier  in the Einstein frame (\ref{1.6}) can be conformally transformed to a $F(R)$ action in the Jordan frame. It is important to point out that the singularity remains in the transformed frame, where  appears as an initial singularity.

\section{Discussions}

In the present paper, we have studied the use of a Lagrange multiplier to reconstruct cosmological solutions, and to study some problems related to  gravitational physics. We have shown that  the usual quintessence/phantom scalar fields,  restricted by the presence of a Lagrange multiplier,  can  reproduce any kind of cosmological solution, while the evolution of the scalar field can be maintained under control, and useful information on the cosmological evolution can be analyzed. In this context, we have proposed a relaxation mechanism of the cosmological constant (cc), which provide a way to study the evolution of the cc value. Hence, a scalar field with no additional propagating wave modes, and whose equations are simpler than the usual case can  parametrize and compensate the value of the vacuum energy density in a natural way, providing a method for a better understanding of the cc problem, whose cosmological evolution can be easily studied. In this sense, we have provided several examples in order to show the evolution of the ``dark energy'' density, and how the value of the vacuum energy density is reduced. The next step will be a deeper study of this kind of models, and its constraints with the observational data, but this is beyond of the purpose of this paper.  In addition,  we have shown that solutions  capable to unify  inflation and late-time acceleration under the same mechanism can be  reconstructed. Naturally this approach is extended to $F(R)$ gravity, where the restriction affects to the metric tensor. Such constraint is expected to become important at large scales, while at small scales the usual equations of General Relativity should be  recovered.  As in the case of a scalar field, the Lagrange multiplier is a powerful tool to control the problems of a given model, and it  gives important information on the evolution of the solutions. \\

On the other hand, the formalism has been also used to study the behavior of future singularities, specifically the so-called Big Rip singularity, when a conformal transformation that led to a type of $F(R)$ gravity in the Jordan frame is applied. It is well known that the reconstruction of $F(R)$ gravity is technically not possible in the usual case of phantom scalar fields. However, in this case, we can have a positive kinetic term of the scalar field defined in the Einstein frame due to the constraint equation provided by the Lagrange multiplier, which reproduce the phantom evolution in the original frame. Hence, this allows to study the behavior of future singularities when this kind of conformal transformation is applied. We have shown that the singularity in the Einstein frame remains also in the Jordan one, but with a different behavior. By allowing the time coordinate  scale with the conformal transformation, we have shown that the Big Rip singularity is transformed into a Big Bang (Crunch) singularity. This may help for a better understanding of the physical relation between both frames, however such discussion is beyond the purpose of this paper (see Ref.~\cite{Faraoni:1998qx}). Other aspects about the relation between both conformal frames and its (un)physical equivalence are recently studied in Ref.~\cite{Capozziello:2011wg}. \\
 
Hence, the use of Lagrange multipliers provides a scalar field model with simpler equations and without additional wave modes, that can be used to study the evolution of the cc value and provide a way to analyze its evolution, and reconstruct the entire evolution of the Universe in a natural way. In addition, we have shown that $F(R)$ gravity can be restored from a model containing a Big Rip singularity, which  gives a better understanding on how future singularities behave. 

\section*{Acknowledgments}
I thank the referee of a previous version of this paper for comments and criticisms that led to its improvement. This work was done during a stay at the California Institute of Technology  (Pasadena, USA), and I would like to thank Sergi Hildebrandt, Jamie Bock, and the  Cahill Center for Astronomy and Astrophysics  for their kind hospitality. I acknowledge a FPI fellowship from MICINN (Spain), project FIS2006-02842.

\end{document}